\documentclass{rstransa}

\newcommand*\apj{Astrophys. J.}
\newcommand*\apjl{Astrophys. J. Lett.}
\newcommand*\aap{Astron. Astrophys.}
\newcommand*\mnras{Mon. Not. R. Astron. Soc.}
\newcommand*\memsai{Mem. Soc. Astron. Ital.}

\hypersetup{
	colorlinks=true,      
	linkcolor=red,        
	citecolor=blue,       
	filecolor=magenta,    
	urlcolor=cyan,   
	pdftitle={On the influence of magnetic topology on the propagation of internal 
		gravity waves in the solar atmosphere},
	pdfauthor={Vigeesh}, 
	pdfsubject={Astrophysics},   
	pdfkeywords={Magnetohydrodynamics (MHD), Sun: atmosphere, Sun: chromosphere, Sun: granulation, Sun: magnetic fields, Sun: photosphere, waves, internal gravity waves}
}

\begin{document}

\title{On the influence of magnetic topology on the propagation of internal 
	gravity waves in the solar atmosphere}

\author{
G. Vigeesh$^{1}$, M. Roth$^{1}$, O. Steiner$^{1,2}$ and B. Fleck$^{\rm 3}$}

\address{$^{1}$Leibniz-Institut f\"{u}r Sonnenphysik (KIS), Sch\"{o}neckstra\ss{}e 6, 79104 Freiburg, Germany\\
$^{2}$Istituto Ricerche Solari Locarno (IRSOL), Via Patocchi 57, 6605 Locarno-Monti, Switzerland\\
$^{3}$ESA Science and Operations Department, c/o NASA/GSFC Code 671, Greenbelt, MD, USA}

\subject{astrophysics, computer modelling and simulation, wave motion}


\corres{G. Vigeesh\\
\email{vigeesh@leibniz-kis.de}}

\begin{abstract}
The solar surface is a continuous source of internal gravity waves
(IGWs). IGWs are believed to supply the bulk of the wave energy for
the lower solar atmosphere, but their existence and role for the
energy balance of the upper layers is still unclear, largely due to
the lack of knowledge about the influence of the Sun's magnetic fields on
their propagation. In this work, we look at naturally excited IGWs in
realistic models of the solar atmosphere and study the effect of
different magnetic field topographies on their propagation. We carry
out radiation-magnetohydrodynamic (R-MHD) simulations of a magnetic
field free and two magnetic models -- one with an initial,
homogeneous, vertical field of 100\,G magnetic flux density and one with an initial
horizontal field of 100\,G flux density. The propagation
properties of IGWs are studied by examining the phase-difference and coherence spectra in the $k_{h}$-$\omega$ diagnostic diagram. We find
that IGWs in the upper solar atmosphere show upward propagation in the
model with predominantly horizontal field similar to the model without
magnetic field. In contrast to that the model with predominantly
vertical fields show downward propagation. This crucial difference in
the propagation direction is also revealed in the difference in energy transported
by waves for heights below 0.8\,Mm. Higher up, the propagation properties show a peculiar behaviour, which require further study.
Our
analysis suggests that IGWs may play a significant role in the heating
of the chromospheric layers of the internetwork region where horizontal fields are thought to be prevalent.
\end{abstract}



\maketitle

\section{Introduction}\label{s:intro}
Waves are a ubiquitous phenomenon in the atmosphere of a star,
including that of the Sun. Of the different types of waves present in
the solar atmosphere, internal gravity waves (IGWs) are perhaps the
least studied of them all. Propagating with frequencies below the
acoustic waves, IGWs  are buoyancy-driven waves naturally occurring in
a continuously stratified fluid. The solar atmosphere happens to be an
ideal environment for their generation, sustenance and eventual
dissipation.

The main driver of IGWs in the solar atmosphere are the convective
updrafts penetrating into the stably stratified atmospheric layer
above. IGWs have been detected in the solar atmosphere and are thought
to significantly contribute towards the total wave energy flux in the
lower atmosphere 
\cite{2008ApJ...681L.125S}. 
Low-frequency ultraviolet (UV) brightness fluctuations observed in the
internetwork region of the low solar chromosphere are believed to be
caused by IGWs dissipating in the higher layers 
\cite{2003A&A...407..735R}.
However, 
the strong effects of magnetic field orientation (attack angle) on the propagation of IGW have been demonstrated by
Newington \& Cally 
\cite{2010MNRAS.402..386N, 2011MNRAS.417.1162N} and
realistic numerical simulations have shown that the magnetic
field present in the solar atmosphere influence the propagation of
these waves and likely hinder them from propagating into the upper
atmosphere 
\cite{2017ApJ...835..148V,2019ApJ...872..166V,2020A&A...633A.140V}. 
At this point it is still unclear what role the magnetic field plays
in the propagation of IGWs into higher layers and how this fits with
the brightness fluctuations seen in chromosphere.

While it is well known that the quiet solar atmosphere is permeated by
magnetic field, the exact topology of the magnetic field depends on
whether one examines a network or an internetwork region on the solar
surface. The network field is mainly characterised by strong vertical
component that form magnetic flux concentrations which merge with
other network field bundles outlining a supergranulation cell. The
area surrounded by the supergranular network (i.e., the cell interior)
is usually referred to as the internetwork, where the magnetic field
topology is of more mixed orientation, often with a predominance of
horizontal fields. The internetwork usually has fewer strong, vertical
flux tubes and is mainly pervaded by weaker horizontal fields. For a
recent review on the observational aspects of the quiet solar
magnetism, the reader is referred to 
Bellot Rubio et al.\cite{2019LRSP...16....1B}.

What does this mean for IGWs propagating in a magnetic environment as
diverse as the solar atmosphere? The combined effect of buoyancy, gas
pressure, and magnetic field results in a more complicated picture of
magneto-acoustic-gravity waves than the one simply described by the
acoustic-gravity wave spectrum, wherein the waves are clearly
decoupled throughout the atmosphere. The complexity arises due to the
additional anisotropy introduced by the magnetic field direction,
leading to a multitude of wave modes that depend locally on the angle
between the direction of gravitational acceleration and the magnetic
field. Furthermore, the highly dynamic and inhomogeneous atmosphere of
the Sun provides an environment in which changes happen rapidly from
region to region rendering it practically impossible to assign an
average property to the atmosphere as seen by a passing wave. This is
much more important for IGWs, as local changes may happen faster than
the characteristic time scales of the waves. 

The effect of magnetic field on the propagation of IGWs in a model
solar atmosphere permeated by a uniform static magnetic field of
different orientations were undertaken by Newington \& Cally
\cite{2010MNRAS.402..386N, 2011MNRAS.417.1162N}. 
They showed that in regions of highly inclined magnetic fields, IGWs
undergo mode conversion to upward propagating (field-guided) acoustic
or Alfv\'{e}nic waves. They suggest that the upward propagating
acoustic waves are then likely to dissipate by shock formation before
reaching the upper chromosphere, while converted Alfv\'{e}nic waves
can propagate to higher layers. In contrast to the case of highly
inclined field, the presence of a vertical field results in the
reflection of these waves back into the lower atmosphere. Radiative
damping effects do not play a significant role in the mode conversion
higher up, but may be important in the surface layers where the IGWs
are likely generated. 

In this paper, we address this problem with more realistic models of
the solar magnetic environment that mimic the atmospheric conditions
as closely as possible. Our aim is to better understand the link
between the brightness fluctuations in the internetwork region and
internal gravity waves, and the role of the magnetic field topology on
their propagation characteristics. This paper extends on the previous
studies 
\cite{2017ApJ...835..148V,2019ApJ...872..166V,2020A&A...633A.140V} 
by comparing the propagation of IGWs in models with different magnetic
field orientations. These environments are far from idealistic and
have magnetic field properties that are more representative of the
network/internetwork-like regions of the solar surface.

The outline of this paper is as follows: in 
\S\ref{s:method} 
we present the models that we use in this study and describe the
data analysis method for investigating waves, in
\S\ref{s:results} 
we present the results of our wave analysis and in 
\S\ref{s:conclusion} 
we provide our conclusions.

\section{Method}\label{s:method}
We carry out full three-dimensional simulations of the near-surface
layer of the Sun using the {CO$^{\rm 5}$BOLD} code 
\cite{2012JCoPh.231..919F}.  
The numerical code solves the time-dependent nonlinear MHD equations
in a 3D Cartesian domain with an external gravity field and including
non-grey radiative transfer. Approximately, the lower half of the
computational domain is in the convective layer and the upper half
extends into the stable atmospheric layer where the waves we are
interested in propagate. The waves are studied by looking at
characteristic properties revealed by their dispersion relation. In
the following, we describe the construction of the models and the data
analysis in more detail.

\subsection{Numerical models}\label{s:method:models}
We consider three models that mainly differ in the initial magnetic
field introduced before the start of the simulation. Firstly, a
non-magnetic model (Sun-v0) is constructed by setting the initial
magnetic flux density to zero. In order to mimic a magnetic
network-like environment, we construct a second model with an initial,
vertical, homogeneous field of 100~G flux density (Sun-v100). Finally,
to represent the magnetic field behaviour in an internetwork-like
region, we build a third model with an initial, horizontal,
homogeneous field of 100~G flux density (Sun-h100). The former two
models were part of an earlier study 
\cite{2019ApJ...872..166V,2020A&A...633A.140V}. 
In this work, we present the new model that is constructed by
introducing a horizontal magnetic field aligned in the $x$ coordinate
direction. 
All models, including the field free model, were calculated with the very same MHD-solver (HLL-MHD;  see Freytag et al. \cite{2012JCoPh.231..919F}).
All three simulations are carried out on a
computational domain of 
38.4$\times$38.4$\times$2.8~Mm$^{\rm 3}$, 
discretized onto a 
480$\times$480$\times$120 mesh. 
The domain extends $\sim$1.3~Mm above and $\sim$1.5~Mm below the mean
Rosseland optical depth $\tau_{R}$=1. For more details on the
numerical setup, the reader is referred to 
Vigeesh et al.
\cite{2019ApJ...872..166V}.

The main difference between the three models is in the initial
condition. Additionally, they also differ in the
prescription of the top and bottom boundary conditions for the
magnetic field. We use periodic boundary conditions for the side
boundaries for all the three models. This dictates the fluid flow,
radiation, and the magnetic field components to be periodic in the
lateral direction. The top boundary is open for flow and radiation and
the in-flowing material at the bottom boundary carries with it a
constant specific entropy to maintain the radiative flux corresponding
to an effective temperature ($T_{\rm{eff}}$) of $\sim$5770~K. For the
vertical field model (Sun-v100), the top and bottom boundary condition
is such that the vertical component of the magnetic field is constant
across the boundary and the transverse component drops to zero at the
boundary. For the horizontal field case (Sun-h100), the vertical
component of the magnetic field is fixed at its initial value of zero
and a constant extrapolation applies to the transverse component
across the boundary. The bottom boundary condition for the Sun-v100
model is the same as its top boundary condition. For the Sun-h100
case, however, the up-flowing material carries with it horizontal
magnetic flux density of 100~G when ascending into the computational
domain.

\begin{figure}[!h]
\centering
\includegraphics[width=1\linewidth]{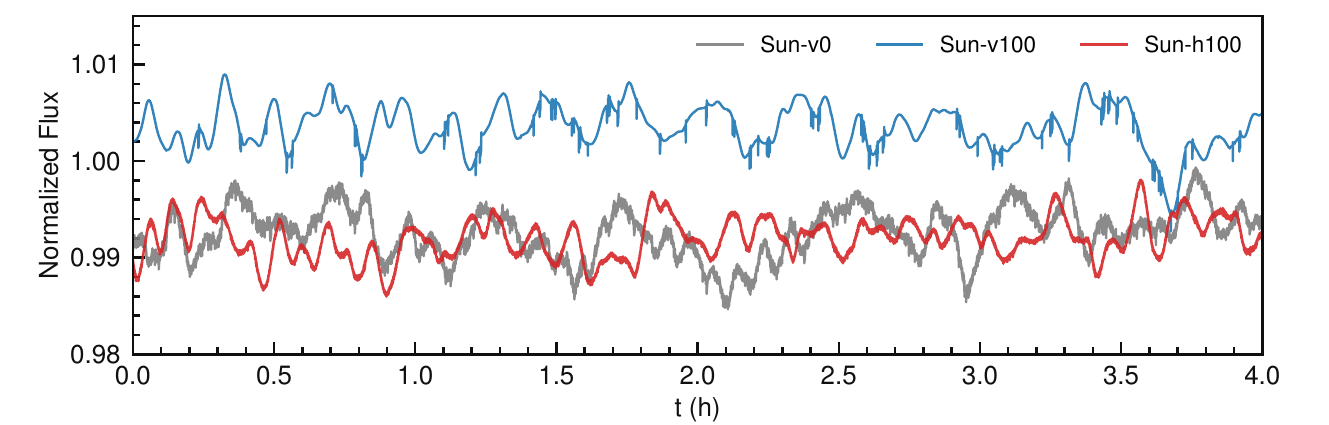}
\caption{Outward radiative flux, normalized by $\sigma
	T_{\rm{eff}}^4$, as a function of time for the non-magnetic (Sun-v0)
	in gray, vertical (Sun-v100) in blue, and horizontal field (Sun-h100)
	model in red.}
\label{fig:frtop}
\end{figure}

The three set of simulations were run for 4~hr with a snapshot
captured every 30~s, giving a total of 480 time points.
Figure~\ref{fig:frtop} shows the outward radiative flux, normalized by
$\sigma T_{\rm{eff}}^4$, as a function of time for the three
simulations. It should be noted that this quantity is measured at a
higher cadence. The vertical magnetic model (Sun-v100) shows a
slightly elevated effective temperature presumably due to the
radiative channeling effect of vertical flux concentration 
\cite{2018A&A...614A..78S}. 
Independent of the different initial and boundary conditions, all models show a stable radiative output throughout the entire span of the simulation run. Having such a similar, stable and continuous time series allows
us to perform Fourier analysis for studying waves and comparing them
on an equal footing.

\begin{figure}[!h]
\centering
\includegraphics[width=1\linewidth]{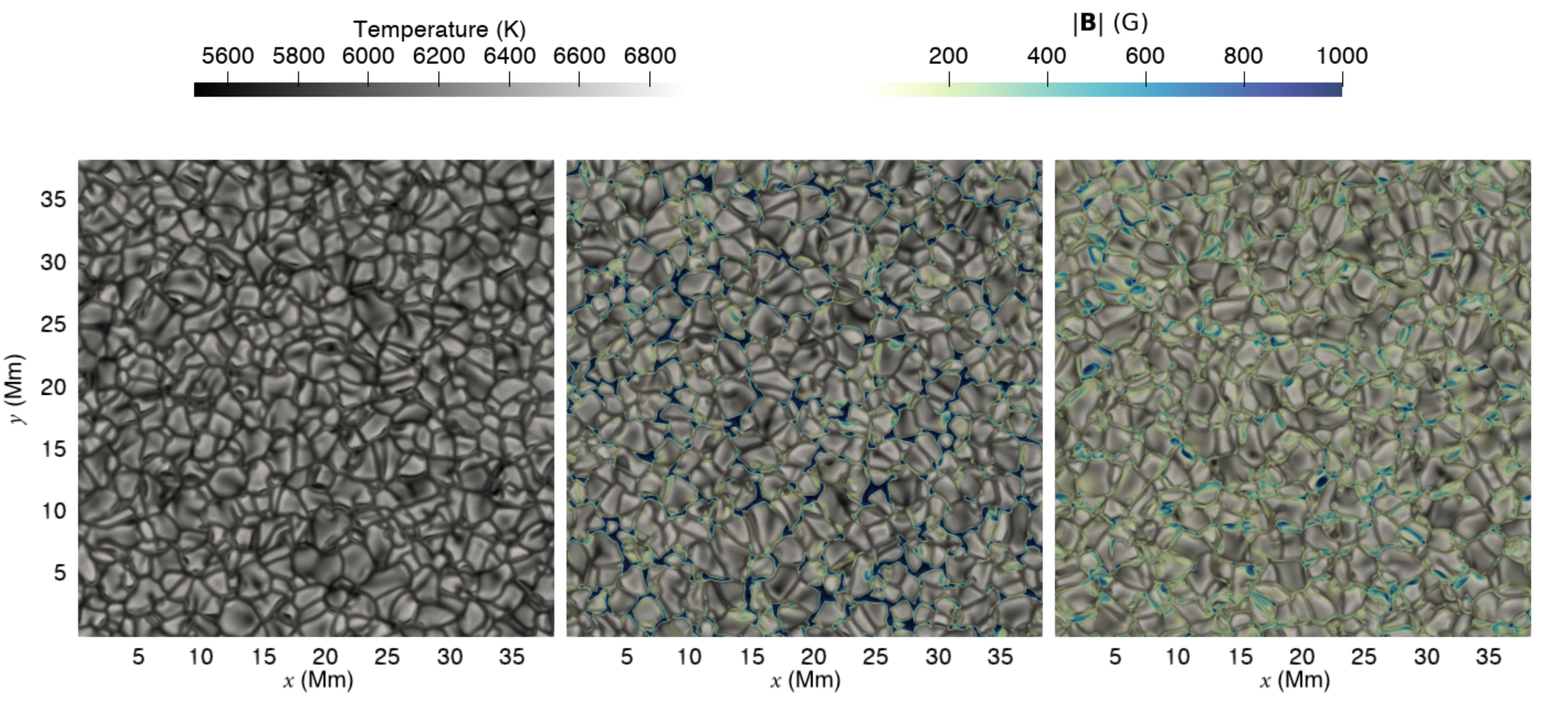}
\caption{Temperature at a reference level of $\tau_R=1$ (in gray) and
	absolute magnetic field strength at $\tau_{R}=1$ (in color with
	$\alpha$-blending to highlight the stronger fields) from the three
	models of solar magnetoconvection: Sun-v0 (left), Sun-v100 (middle),
	and Sun-h100 (right). The snapshots shown here are taken 3 hr after
	the start of the simulation.}
\label{fig:temperature}
\end{figure}

Figure \ref{fig:temperature} shows the temperature at a reference
level of $\tau_R=1$ and the absolute magnetic field strength at the
same layer, for snapshots taken 3 hours after the start of the
simulation for each of the three models. The granular structure,
represented here in grayscale, is similar in all the models.  The
total magnetic field strength is overplotted on the two magnetic
models: Sun-v100 (center) and Sun-h100 (right). We note from these
maps that most of the magnetic flux in the Sun-v100 is preferentially
located in the intergranular lanes, while the Sun-h100 model does not
harbor as much vertical flux concentrations as the former case. The
magnetic field in the Sun-h100 case is more scattered and shows mixed
orientations at the $\tau_{_R}=1$ reference layer.

\begin{figure}[!h]
\centering
\includegraphics[width=1\linewidth]{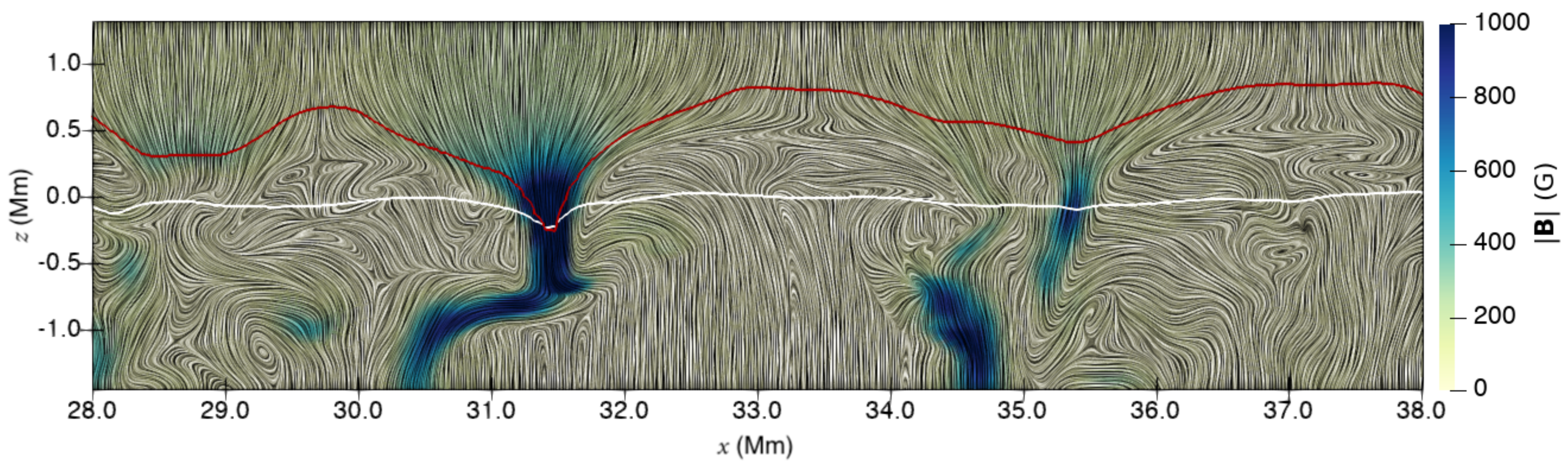}\\
\includegraphics[width=1\linewidth]{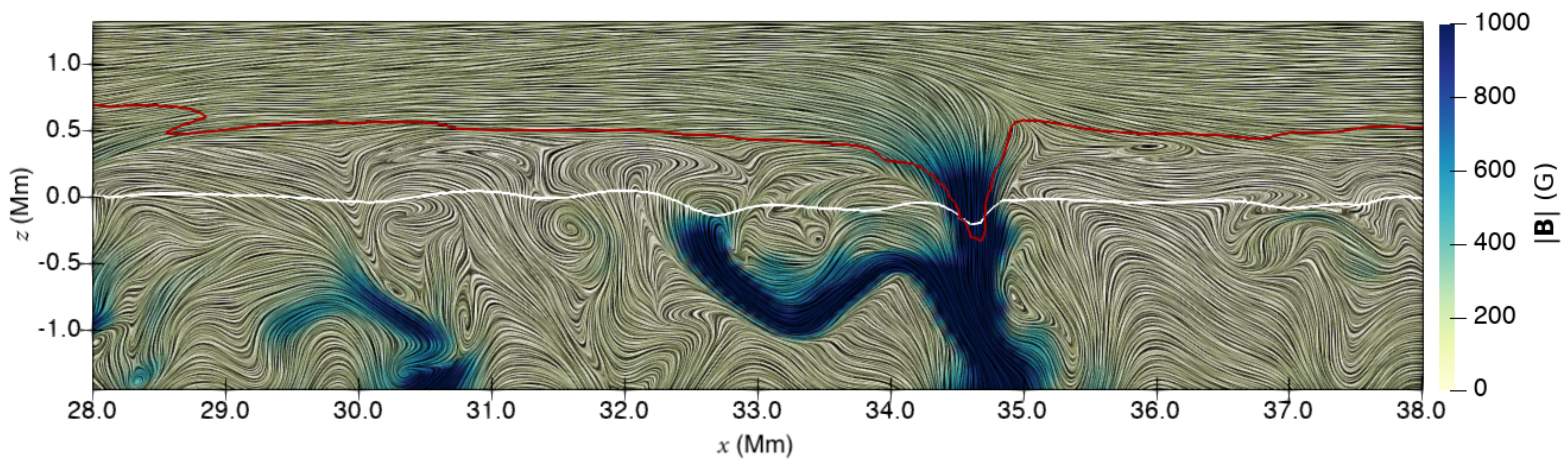}
\caption{Visualization of magnetic fields (colored) in the
	{${x}$-${z}$} plane showing the predominant orientation of the field
	in the two magnetic simulations: Sun-v100 (top) and Sun-h100 (bottom).
	The white contours denote the $\tau_{_R}$=1 surface and the red
	contour is the plasma-$\beta$=1 surface.}
\label{fig:magnetic_fields}
\end{figure}
The difference in topology of the magnetic models in both cases can be
seen by examining a vertical section through the computational domain.
In Figure~\ref{fig:magnetic_fields} we show a representation of the
magnetic field lines on a $x$-$z$ plane in a small region of the
Sun-v100 (top panel) and Sun-h100 (bottom panel) models. The region is
arbitrarily chosen to show a fully developed magnetic flux
concentration in both of the models. Also shown overlaid on the plots
are the contours of $\tau_{_R}$=1 in white and the plasma-$\beta$=1 in
red, where plasma-$\beta$ is the ratio of the gas pressure to magnetic
pressure. The difference between the two models can be clearly
discerned from the magnetic field orientation. The main difference
occurs above the equipartition surface $\beta$=1 layer, where for the
Sun-v100 model we have predominantly vertical field components and for
the Sun-h100 model the field lines are predominately horizontal with
vertical footpoints connecting them to the intergranular lane. The
plasma-$\beta$ contour in these two examples is shown to dip below the
$\tau_{_R}$=1 contour with the magnetic field strength surpassing 1~kG
at these locations. Between the height where the average $\tau_{_R}$=1
and average plasma-$\beta$=1, the two models are similar and show
mixed orientation of the magnetic field but with a preference to their
initial direction, except of course at the locations of flux
concentration where they are vertically aligned.

\begin{figure}[!h]       
	\centering\includegraphics[width=.47\linewidth]{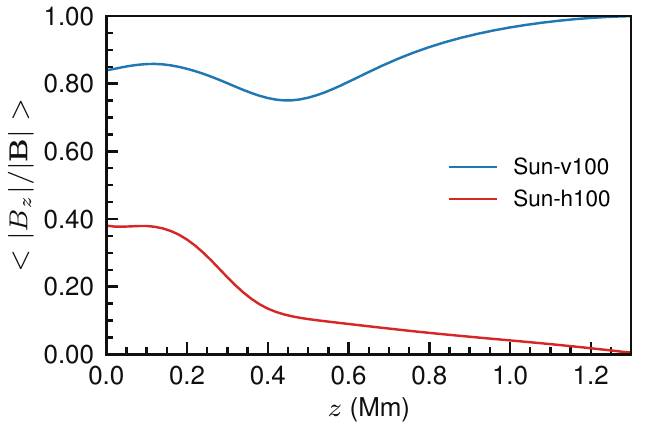}\hskip3ex\includegraphics[width=.47\linewidth]{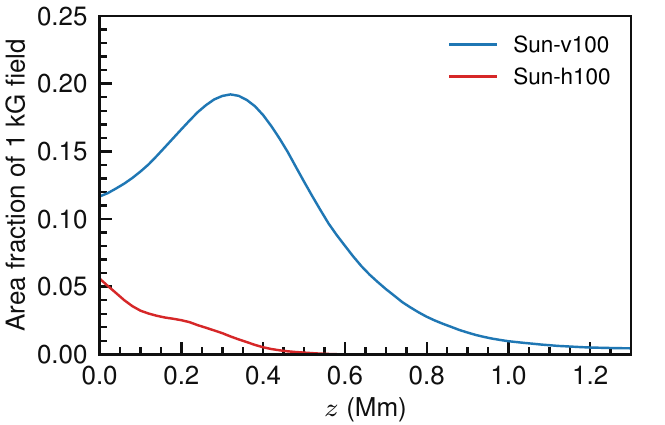}
	\caption{Ratio of mean absolute vertical component of magnetic field ($B_{z}$) to the total mean $\mathbf{B}$ (left panel) and the area fraction covered by $|\mathbf{B}|>1$ \,kG field (right panel) in the Sun-v100 (blue) and Sun-h100 (red) models as a function of height.}
	\label{fig:flux_full_factor}
\end{figure}

In Figure~\ref{fig:flux_full_factor}, the left panel shows the ratio of mean absolute vertical component of the magnetic field ($|B_{z}|$) to the absolute field strength $|\mathbf{B}|$ of both models and the right panel shows the area fraction covered by $|\mathbf{B}|>1$ \,kG field of both models as a function of height. In the case of Sun-v100, for layers above $z=0$\,Mm, we see that a fraction of more than 0.75 of the magnetic flux density is vertically directed and increases with height above $z=0.5$\,Mm reaching 100\% at the top boundary due to the boundary condition. The area fraction that is covered with $|\mathbf{B}|>1$\,kG in the Sun-v100 model is 0.1 -- 0.2 in the range $z=0$ to 0.5\,Mm, decreasing with height further above. The initial increase is the result of the fanning of magnetic flux concentrations in the lower part of the atmosphere. In the case of Sun-h100, only less than 40\% of the magnetic flux density is vertically directed in the lower solar atmosphere, decreasing with height to 0\% at the top boundary as a result of the boundary condition. The area fraction that is covered with $|\mathbf{B}|>1$ \,kG in the Sun-h100 model is less than 0.06.

Measurement of the quiet Sun magnetic fields is particularly difficult in the upper solar atmosphere.
Indirect methods using spectropolarimetric data reveal a prevalence of horizontal fields in the internetwork region 
\cite{2008ApJ...672.1237L}, with loop like structures portruding from the surface
\cite{2010ApJ...714L..94M}. 
These loops show a ``flattened'' geometry as they rise through the temperature minimum region. We believe that the internetwork region is pervaded with such loops and in the vicinity of the temperature minimum region they might tend to have a predominantly horizontal field topology due to their ``flattened'' nature. The network region on the other hand can be thought of as being dominated by the footpoints of the low-lying loops as well as long reaching vertical flux concentrations, thereby giving them a predominantly vertical field topology.

\subsection{Wave diagnostics}
After a sufficient duration of the simulation is complete, 4 hrs in
our case, we extract different physical quantities from the simulation
to perform a spectral analysis. The spectral analysis is carried out
by decomposing the physical quantities into their Fourier components
in the horizontal direction and in time for each grid level in the
$z$-direction. The Fourier synthesis is performed using the fast
Fourier transform (FFT) algorithm, and the components are then
represented on a {$k_{h}$-$\omega$} dispersion relation diagram for a
given height by azimuthally averaging over the $k_{x}$-$k_{y}$ plane.
The phase difference and coherence spectra of the velocities at two
separate layers are then computed from the complex cross-spectrum, 
$\displaystyle S_{v_1,v_2}(\boldsymbol{k},\omega)= v_{1}(\boldsymbol{k},\omega)\cdot \overline{v_{2}(\boldsymbol{k},\omega)}$, 
where $v_{1}$ represents the velocity at layer 1 and likewise for
$v_{2}$, and the overbar represents the complex conjugate  
(see Eqs.~4-6 of 
Vigeesh et al.
\cite{2017ApJ...835..148V}). 
The confidence interval for the phase and coherence measurements and
the zero coherence threshold are also computed as described in 
Vigeesh et al.
\cite{2020A&A...633A.140V}. 
To study the energy transport by these waves, we estimate the
mechanical energy flux spectra by computing the perturbed
pressure-velocity ($\Delta p - \boldsymbol{v}$) co-spectrum
represented on the $k_{h}$-$\omega$ dispersion relation diagram.

\section{Results}\label{s:results}
In this section, we present the results from our wave analysis that
show the differences in wave propagation between the three models
introduced in \S\ref{s:method}(a). We then discuss about the
similarities and differences in energy transport by IGWs in the
different models. Lastly, we briefly mention the effect of the
background flow on the dissipation of these waves.

\subsection{Wave generation and propagation}
\begin{figure}[!h]
	\centering\includegraphics[width=.98\linewidth]{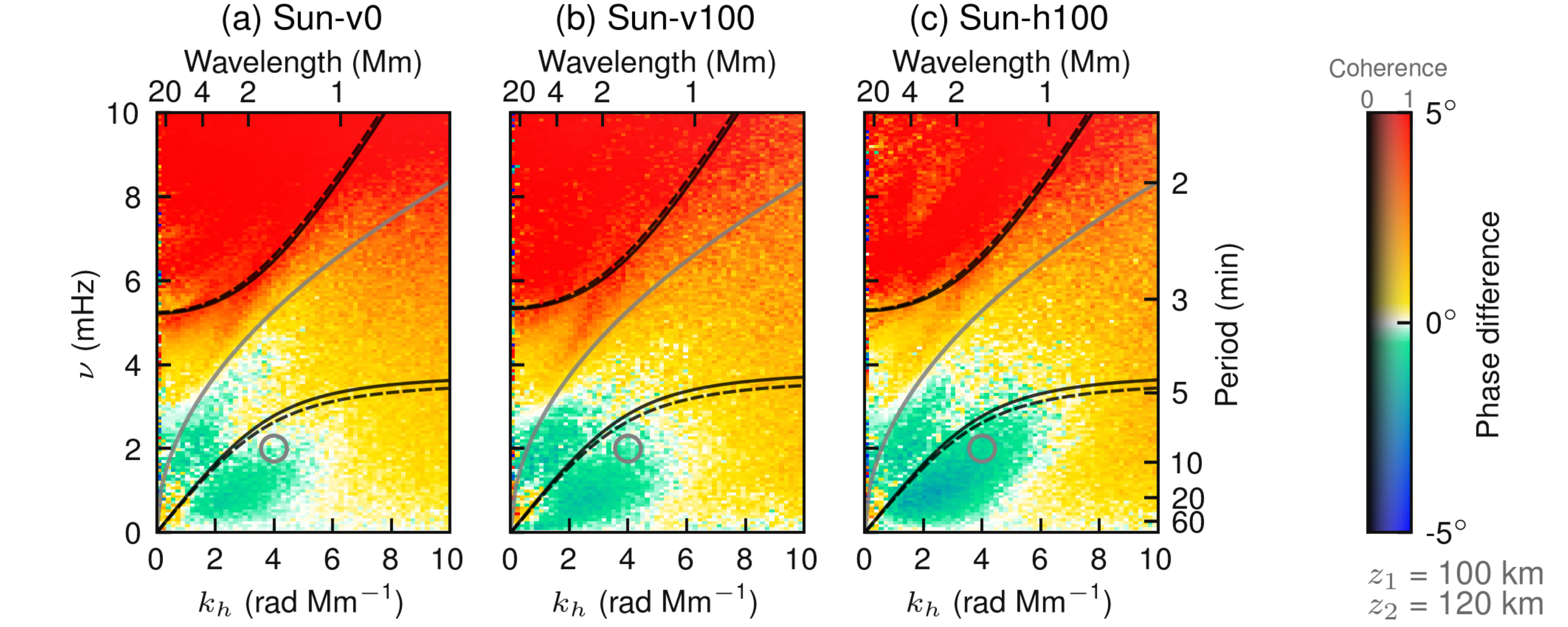}
	\caption{$v_{z} - v_{z}$ phase and coherence spectra estimated between $z=100$\,km and $z=120$\,km for the (a) non-magnetic model, Sun-v0, (b) model with predominantly vertical fields, Sun-v100, and (c) model with the predominantly horizontal fields, Sun-v100. The dashed black curves represent the propagation boundaries for the lower height, and the solid curves represent those for the upper height. The gray curve represents the dispersion relation of the surface gravity waves. The colors represent the phase difference and the shading shows the coherency. IGWs propagate in the region below the lower propagation boundaries.}
	\label{fig:phase_diff_100_120}
\end{figure}

We study the propagation characteristic of the waves by examining the
velocity-velocity ($v$-$v$) phase difference spectra represented in
the $k_{h}$-$\omega$ diagnostic diagram for a given pair of heights. 
Observed $k_{h}$-$\omega$ phase diagrams usually result from time
series of simultaneously acquired Dopplergrams in a number of spectral
lines, formed at different heights. For the simulation, we can extract
the velocity field for any pair of atmospheric layers and compute the
phase-difference spectra between those. In Figure \ref{fig:phase_diff_100_120}, we show the phase difference between the vertical velocity measured at $z=100$\,km and $z=120$\,km of the three models, represented on the  $k_{h}$-$\omega$ diagnostic diagram. This near surface region where the waves are presumably excited show similar IGW emission characteristics -- downward phase propagation (negative phase difference) in the region below the IGW propagation boundary (marked by the lower black solid and dashed curves). The corresponding energy transport is upward directed, indicating that these waves are indeed IGWs. In the rest of the paper, we refer with ``upward'' or ``downward'' propagation to the direction of energy transport, which for gravity waves is opposite to the direction of phase propagation.
An upwardly propagating gravity wave has downwardly propagating phase and shows up with a negative phase difference according to our sign convention when calculating phase difference spectra (phase of lower layer minus phase of upper layer).

\begin{figure}[!h]       
\centering\includegraphics[width=.47\linewidth]{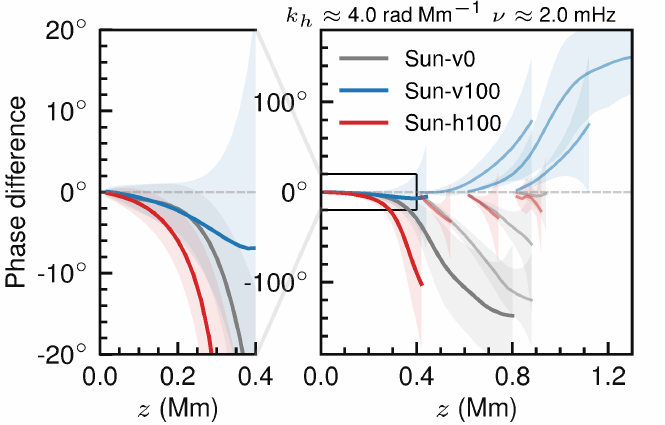}\hskip3ex\includegraphics[width=.47\linewidth]{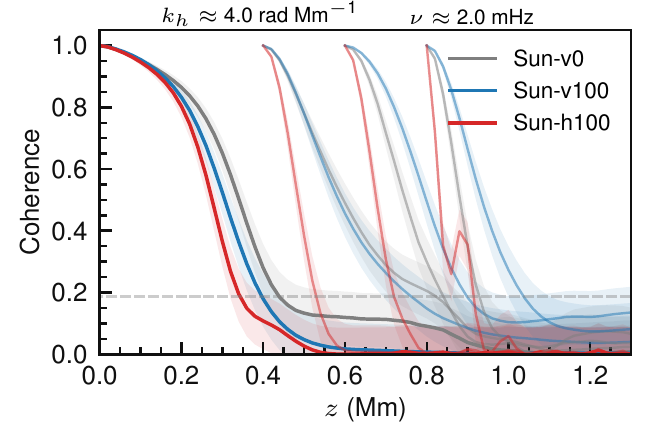}
\caption{\textit{v}-\textit{v} phase difference (left) and coherence
	(right) between the reference layer, $z=0$\,Mm, and layers of
	constant geometrical scale for a given $k_h$ and $\omega$ (thick
	curves) for the three models. The leftmost subplot is a rescaled part
	of the small region in the panel, marked by a black rectangle. The
	thin solid lines show three other reference layers. The 90\%
	confidence bounds for estimates are represented by the shaded area.
	The zero-coherence threshold at a significance level of 0.05 is marked
	by the dashed line in the right plot.}
\label{fig:phase_coherence}
\end{figure}

Although the near surface region show similar wave spectra, higher up in the atmosphere the three models show markedly
	different behaviour on the $k_{h}$-$\omega$ phase difference spectra
	for a given pair of heights.
Rather than limiting ourselves to a
single pair of heights, we look at how the phase difference and
coherence vary as a function of vertical distance between the
measurement heights in the three models. Figure~\ref{fig:phase_coherence} shows the phase difference and
coherence as a function of travel distance relative to a reference
height of $z$=0~Mm (thick curves) for the non-magnetic (grey: Sun-v0),
the vertical field model (blue: Sun-v100), and the horizontal field
model (red: Sun-h100). This is computed by estimating the
$v_{z}$-$v_{z}$ phase difference and coherence between $z$=0~Mm and
every grid point along the $z$-direction. The plots correspond to a
given Fourier component, here $k_{h}\approx 4$~rad~Mm$^{-1}$ and
$\nu\approx 2$~mHz, which fall in the region of the $k_{h}$-$\omega$
where the bulk of the IGWs occur in our simulation (location marked by the circle in Fig~\ref{fig:phase_diff_100_120}). The 90\%
confidence bounds for the phase difference and coherence estimates are
shown by the shaded areas and the zero-coherence threshold is shown by
the dashed-grey line. We also show the phase difference and coherence
relative to a selection of other reference layers, viz. $z$ = 0.4,
0.6, 0.8 Mm. These plots clearly reveal the difference in the
propagation properties of waves in models without, with vertical, and
with horizontal magnetic fields.

Firstly, all the three models show negative phase difference up to
around 0.4~Mm, suggesting that the waves in these model propagate in
the upward direction (downward
propagating phase). We note here that beyond the travel distance of
0.4~Mm for the $z=0$~Mm reference layer, the phase difference
measurements in both the magnetic models become unreliable as
indicated by the break in the phase-difference spectra and by the
uncertainty in the measurement shown by the shaded area. The coherence
in all the three models above a height of 0.4~Mm drops below the
zero-coherence threshold. 

Particularly interesting to note is the difference in the phase and
coherence spectrum for the reference layer of $z=0.4$~Mm and the same
above. Here, we see that the non-magnetic model always shows negative
phase-difference, suggesting that the waves are upward propagating
throughout the atmosphere. However, when comparing the two magnetic
models, we clearly see the influence of magnetic field orientation on
the propagation of the IGWs. The model with a predominantly horizontal
field reveals a similar behaviour as the non-magnetic case, showing
negative phase-difference throughout the atmosphere, meaning that the
IGWs propagate upwards as if the magnetic fields were absent. For the
model with a predominantly vertical field, on the other hand, this is
not the case as the phase difference here is positive for heights
above $z=0.4$~Mm, revealing quite a contrasting behaviour compared to
the non-magnetic and the model with horizontal field. The IGWs in the
vertical field case are seen to propagate downwards in the higher
layers.

The change in behaviour around $z=0.4$~Mm can be understood by looking
again at Figure~\ref{fig:magnetic_fields}. We see here that the
magnetic fields in the two models are quite similar in regions between
the average $\tau_{_R}=1$ surface and average $\beta=1$ surface which
is around $z=0.5$~Mm. 
We believe that the similarity in both simulations in terms of wave propagation is due to the fact that the dynamics of the waves below the plasma $\beta$=1 surface is not dominated by the magnetic fields, but by the thermodynamic properties of the atmosphere. The thermal properties of the atmosphere in the near surface layers are similar in the two cases of Sun-v100 and Sun-h100.
In the low plasma $\beta$ layers, however, the waves see a different
magnetic field orientation between the two magnetic models as they
propagate. This is an indication of the mode
coupling that occurs in the presence of magnetic fields of different
inclination as described by 
Newington \& Cally
\cite{2010MNRAS.402..386N, 2011MNRAS.417.1162N}.

\subsection{Energy transport}
The difference in the wave propagation behavior as revealed by the
phase and coherence analysis presented in the previous section
highlights an important aspect of the energy transport of these waves
in the solar atmosphere. 

\begin{figure}[!h] 
\centering \includegraphics[width=0.5\linewidth]{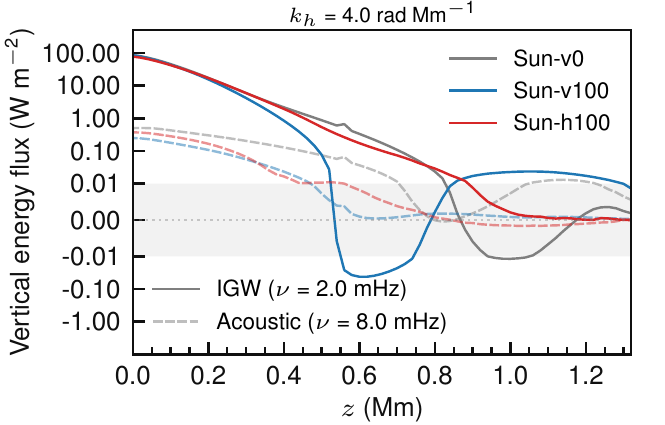}
\caption{Mechanical flux as a function of height for a given $k_h$ and
	$\omega$ for the different models.}
\label{fig:energy_flux}
\end{figure}

In order to better understand the energy transport of the waves in the
different models, we look at the mechanical energy flux computed in
the $k_{h}$-$\omega$ diagnostic diagram for a given layer. In
Figure~\ref{fig:energy_flux}, we show the mechanical flux as a
function of height for the three models. Here, we show it for the same
Fourier component ($k_{h}\approx 4$~rad~Mm$^{-1}$, $\nu\approx 2$~mHz)
as for the phase/coherence analysis above. In addition, for comparison
we also show the fluxes for a Fourier component that correspond to the
high-frequency acoustic wave ($k_{h}\approx 4$~rad~Mm$^{-1}$,
$\nu\approx 8$~mHz).

We see that for all the three models, the mechanical energy flux is
upward directed up to a height of $z\approx0.5$~Mm, confirming the
propagation properties that are revealed by the phase-difference
analysis. Above this height, we see that the non-magnetic model and
the model with horizontal field show a very similar behaviour
exhibiting upward energy transport up to a height of $\sim$0.9~Mm,
beyond which they tend to deviate. The reason for this behaviour is
unclear. Nevertheless, the plot of the energy flux as a function of
height suggests that the IGWs in a predominantly horizontal field
environment may transport energy to higher layers, up to a height of
$\sim$0.9~Mm. On the other hand, the model with vertical magnetic
field shows a completely different behaviour above $z=0.5$~Mm. Here,
we see that the energy transport is downward directed, which is what
we expected from the phase-difference analysis. This suggests a
markedly different energy transport behaviour of the IGWs in the
presence of vertical magnetic field compared to a non-magnetic
atmosphere or an atmosphere with a predominantly horizontal field. But
here again, the behaviour around $z=0.8$~Mm and above remains unclear. 
We refrain from interpreting the results in the vicinity of the upper boundary ($\sim$ 2-3 pressure scale heights), i.e above $z=1$\,Mm in the atmosphere where the coherence value drops below the zero-coherence threshold at 95\% level anyway.

The peculiar behaviour above the height of $z=0.8$\,Mm in all the three
models suggests that this might have a different origin than the
magnetic field itself.
The measured phase difference and the energy transport for the chosen Fourier component do not fit the behaviour of IGW above $z=0.8$\,Mm, the basic signature of which is the opposite sign of the vertical component of the phase propagation and energy transport. We do not have an explanation for this behavior and these waves might not be IGWs anymore. This is particularly surprising for the Sun-v0 case, because the selected Fourier component falls in the IGW range and should satisfy the acoustic-gravity dispersion relation. The boundary condition seems to have an influence only in the grid cells very close to the boundary. Perhaps a linear analysis breaks down due to the large amplitude of the thermodynamic perturbations in these layers. The temperature perturbations in the upper atmosphere is particularly strong for the Sun-v0 case.
Exploring this aspect is
extremely important, but is beyond the scope of this work. A
cause for the puzzling behaviour above the $z=0.8$~Mm level might be
non-linear effects as is addressed in the following section.

\subsection{Wave-breaking}
When waves propagate in the presence of a background flow, their
propagation properties are modified.  Strong background flow may
result in nonlinear breaking of IGWs, leading to enhanced energy
dissipation. In the context of IGWs in the solar atmosphere, 
\cite{1981ApJ...249..349M} 
considered a stability condition given by the ratio of wave vorticity
($\zeta$) and the Brunt-V\"{a}is\"{a}l\"{a} frequency, $N$, defined
as,

\begin{align}\label{1.1}
N^2 = g \left( \frac{1}{H_{\varrho}} - \frac{1}{\gamma H_{p}}\right),
\end{align}
where, $\gamma$ is the ratio of the specific heats ($c_{P}/c_{V}$),
$H_{\varrho}$ is the density scale height, and $H_{p}$ is the pressure
scale height of the atmosphere.

\begin{figure}[!h]
\centering\includegraphics[width=.47\linewidth]{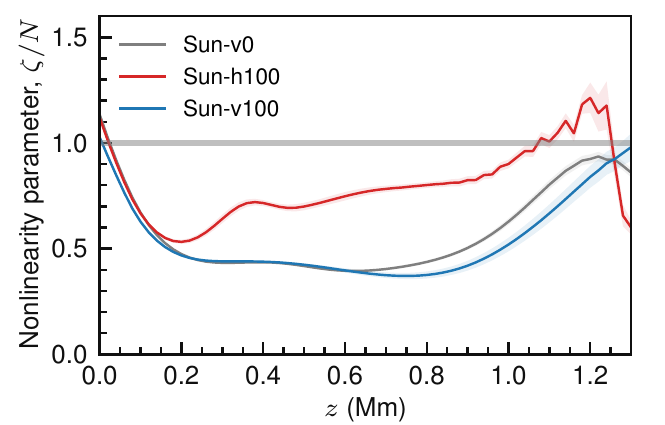}\hskip3ex\includegraphics[width=.47\linewidth]{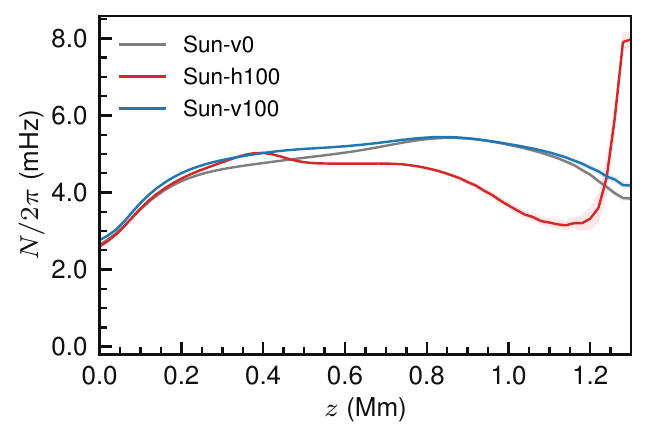}
\caption{Nonlinearity parameter ($\zeta/N$, left panel)  and the Brunt-V\"{a}is\"{a}l\"{a} frequency ($N$, right panel) as a function of height
	in the three models. The peculiarities apparent in the Sun-h100 model above $z\approx1.2$\,Mm are due to the influence of the upper boundary condition.}
\label{fig:nonlinear}
\end{figure}

A non-zero vorticity may occur due to vortex flows or as a result of
velocity shear. Here we consider the average fluid vorticity
($\mathbf{\nabla} \times \mathbf{v}$) as a proxy for the wave
vorticity ($\zeta$) to estimate the strength of the background flow
\cite{2017MmSAI..88...54V}. In Figure~\ref{fig:nonlinear}, the left shows the
ratio, $\zeta/N$ and the right panel shows the  Brunt-V\"{a}is\"{a}l\"{a} frequency as a function of height for the three models. We see
that for most of the atmosphere, $\zeta/N$ is below unity, suggesting
that the model does not have strong background flows to influence the
propagation of IGWs. However, the horizontal field model seems to have
a larger $\zeta/N$ at higher layers compared to both the vertical
field and non-magnetic case, suggesting that IGWs are more likely to
break in the former case. 
The Sun-h100 model shows a lower Brunt-V\"{a}is\"{a}l\"{a} frequency compared to the other models, as a result of the higher density and pressure scale height.
A more detailed analysis would require
estimating the wave vorticity in the {$k_{h}$-$\omega$} diagnostic
diagram and exploring the ratio for a given Fourier component, which
is planned for the future. However, from the energy transport
perspective, we have seen that horizontal magnetic field allow IGWs to
propagate upwards into the atmosphere. Combined with the fact that,
horizontal models also show stronger non-linearity parameter, this suggests that
wave breaking may be a possible scenario by which IGWs dissipate their
energy in an atmosphere with predominantly horizontal fields. Vertical
fields, on the other hand, appear to prohibit IGWs from propagating
into higher layers and therefore prevents them from undergoing
wave-breaking at chromospheric heights.

The numerical solver used in this work is capable of capturing strong discontinuities related to wave breaking condition, but the coarse resolution of the simulation
restricts us from exploring this aspect adequately.	
The models presented in this work are of low resolution ($\delta x, \delta y = 80$\,km) that do not capture strong vorticity, and therefore it is unclear if wave breaking is still the reason for the decrease in energy flux that we observe. 
We see that the average temperature remains similar in the three models. Therefore, we cannot claim to have clear evidence of wave breaking or the consequent temperature enhancement in our simulation.	
High resolution simulations show significantly stronger flows and vorticity 
\cite{2017MmSAI..88...54V, 2020arXiv200705847F,2020A&A...639A.118C}
and therefore maybe better suited for the study of wave breaking.

\section{Conclusion}\label{s:conclusion}
The complex and highly dynamic atmosphere of the Sun harbours internal
gravity waves. They have been detected in observations and have been
clearly reproduced in realistic three-dimensional numerical
simulations of the solar atmosphere. In this work, we show that the
energy flux spectra in the lower photosphere are dominated by upward
propagating IGWs and are independent of the magnetic properties of the
model.  In the higher layers of the atmosphere, the average magnetic
field orientation dictates the propagation properties of the IGWs. In
vertical field models the IGWs are reflected downwards, whereas in
models without a magnetic field and a horizontal magnetic field they
propagate freely into the upper layers. Our study demonstrates that
IGWs behave similarly in the near surface layers of internetwork-like
region where the fields are predominantly horizontal and in network
regions where the fields are predominantly vertical. In these low
layers, the magnetic field does not play a significant role. However,
the significant differences in behaviour between the two models in the
upper layers above $\approx$ 0.4 Mm, demonstrate the importance of the
orientation of the magnetic field for the propagation of IGWs. The
upward propagation of IGWs in internetwork-like regions may lead to
wave-breaking and therefore may be related to the brightness
fluctuations seen in UV passbands. This is in agreement with the interpretation of the observed UV brightness fluctuations as signature of IGWs by Rutten et al.
\cite{2003A&A...407..735R}. 
On the other hand, where there are
predominantly vertical field components, like in network-regions, IGWs
tend to reflect back without reaching wave-breaking heights and
therefore may play only a minor role for the heating of the upper
layers. 
The peculiar behaviour of the waves above $z=0.8$\,Mm in our simulation is not understood and requires further study.
A detailed comparison with observations should shed further
light on the differences in wave propagation behaviour between
internetwork and network regions.

\vskip6pt

\enlargethispage{20pt}


\dataccess{The data this study is based on are too large to host on
	public repositories. However, parts of the data can be requested from
	the corresponding author, who will be happy to discuss ways to access
	the data.}

\aucontribute{GV designed and carried out the simulations, performed
	the data analysis and drafted the manuscript.  All authors read and
	contributed to the discussion and helped polish the manuscript.}

\competing{The author(s) declare that they have no competing interests.}

\funding{This work was supported by the
\emph{Deut\-sche For\-schungs\-ge\-mein\-schaft, DFG\/} grant RO 3010/3-1.}

\ack{We thank the anonymous referee for exceptionally detailed and constructive comments, which helped to significantly improve the paper.}



\end{document}